\documentclass[12pt]{article}%
\usepackage{ amsmath, graphics, rotating}%

\begin{document}
\large
\begin{center} {\bf Tensor Potential Description of Matter and Space} \end{center}
\begin{center}
{Boris Hikin}
\small

E-mail:  boris.hikin@verizon.net

Tel: 310-922-4752, or 310-826-0209, USA \end{center}
\vskip 1em
\large
{\it Abstract:}
\vskip 0.5em
\small
A unified field theory for the description of matter in a curved space is discussed. 
The description is based on a standard Lagrangianian formalism in a pseudo-Euclidian 4D continuum
using a 3-index tensor as independent variables.
The theory defines gravitational matter as a vector field and contains the Einstein-Schwarzschild solution for 
the metric tensor in the case of stationary spherical symmetry. 
Maxwell's equations, generalized for curved space, offer an alternative to the expanding Universe explanation
for the red shift phenomenon.
The theory contains a covariant law for conservation of matter and strictly supports the Mach principle.
A perturbation method allows a natural transition to the special relativity, providing an explanation and
the scale of elementary particles parameters. 

\baselineskip 1pc
\parindent 2cm
\large
\vskip 3em
{\bf \underline {Introduction}}
\vskip 1em

By proposing his theory of General Relativity (GR), in which gravitation is fully
explained from a geometrical point of view, Einstein made the first step in what would become
a 100-year-long march toward the geometrization of physics.
Two common threads can be seen in most geometrization theories: 
a) broadening the definition of geometrical objects - be it affine connections,
the metric tensor, the dimension of space or a combination of the above - to yield,
in addition to the metric tensor $g_{ij}$, the tensor that describes matter;
and b) yielding, in some form, Einstein's equation
\begin{equation}
R_{ij}-\frac{1}{2}Rg_{ij}=T_{ij}
\label{f1}
\end{equation}

The first attempt to geometrize matter came almost immediately after Einstein's work. In 1918, Weyl \cite{i1} offered 
his theory of gravitation-electromagnetism based on widening the definition of affine
connections by adding a vector potential $A_i$:

\begin{equation}
\Gamma^i_{jk}=\gamma^i_{jk}+g^{is}(A_jg_{sk}+A_kg_{sj}-A_sg_{jk}) \nonumber\\
\end{equation} 
Here $\gamma^i_{jk}$ are Christoffel symbols and the Riemann tensor defined as
\begin{equation}
\label{f1a}
R^i_{jkl}={\Gamma}^i_{jl,k}-{\Gamma}^i_{jl,l}+{\Gamma}^i_{mk}{\Gamma}^m_{jl}-{\Gamma}^i_{ml}{\Gamma}^m_{jk}
\end{equation}
serves as the tensor of matter.

Eddington \cite{i2} widened the definition of affine connections even further by postulating that all $\Gamma$s (symmetrical in his case,
$\Gamma^i_{jk}=\Gamma^i_{kj}$) are independent variables in his unified theory and $g_{ij}$ is proportional
to the generalized Ricci tensor. Einstein \cite{i3} extended the definition of $\Gamma$s to an asymmetric 64 component tensor. 
Schrodinger \cite{i4} tried a non-symmetric metric tensor ($g_{ij}\neq g_{ji}$) and used the Palatini variational 
principle to define $\Gamma$s, which in turn defined a generalized Riemann tensor $R^i_{jkl}(\Gamma)$ per eq. (\ref{f1a}).
In the 1950s, there were attempts to view the geometric Riemann tensor
$R_{ijkl}( g_{ij})$ as the tensor of matter - a so called "already unified" theory (\cite{i5}, \cite{i6}).
In the early 1920s, Kaluza \cite{i7} offered a theory of unification of gravitation and electromagnetism based
on a 5-dimensional space.
Kaluza's idea of using multiple dimensions as a basis for the unified theory still dominates theoretical physics today.

These are just few examples \cite{i14}, \cite{i17} of the attempts of the geometrization of physics.
It would not be much of an exaggeration to say that all geometrization theories have met their share of difficulties.

One of the reasons for this lies in the desire to yield the Einstein equation of GR. Even though it is universally accepted, 
GR has its mathematical, physical and philosophical problems. The most important one is
the problem of vacuum. From a philosophical point of view, one would like to think that in the real physical world space is 
a manifestation of matter and cannot exist without matter - Mach principle \cite{i8}, \cite{i9}.

However, GR, as well as Newton's theory, allows for the existence of space without any matter in it.
Interestingly enough, Newton himself did not believe in the idea of vacuum:
"That one body may act upon another at a distance through a vacuum without the mediation of anything else ...
is to me so great an absurdity that I believe that no man, who has in philosophical matters a competent
faculty for thinking, can ever fall into it." (Dicke quoting Newton, \cite{i8})

This difficulty has lead to multiple attempts to add a matter field to GR. 
The most referenced theory belongs to Brans-Dicke \cite{i9}, where gravitational matter is described by the scalar field $\phi$. 
The equations are derived from the Lagrangian L($g_{ij}$, $\phi$ , $\phi_{,j}$). Again, the Lagrangian is chosen
to fit the Einstein equation: $R_{ij}-1/2Rg_{ij}=T_{ij}(\phi, \phi_{,j},g_{ij})$. 
Unfortunately, this theory was not successful in explaining the experimental facts of GR \cite{i10}.

Another difficulty of Einstein's General Relativity is the absence of the law of conservation. 
Einstein's equation $R_{ij}-1/2Rg_{ij}=kT_{ij}$ leads to $T^j_{i;j}=0$, which is viewed  by most as an
absolutely necessary law of physics. And yet, as it is well known, it does not represent conservation
of any particular parameter or value - at least from the point of view of tensor analysis.
This fact forced Einstein to introduce the pseudo-tensor of energy-momentum, which has been strongly criticized
 \cite {i11}.

This paper attempts to put forward a theory that would be free of these dificulties.
The theory considers a more traditional, non-geometric approach for the description of matter and space. 
This approach is a natural extension of the well established "flat space" Lagrangianian field theory,
of which Maxwell's electromagnetic theory is only one example. Instead of the geometrization of matter,
this article proposes an approach that could be viewed as the materialization of space.
The proposed theory contains covariant formalism and
curved space, both of which are essential features of GR.
However, it places  much more emphasis on Mach's principle, in which space can
not exist without matter. This theory defines gravitational matter and shows that it is essential for the formation of space.
It gives an alternative to Einstein's formulation for the laws that govern curved space which are capable of explaining 
the experimental results of Einstein's GR.
The theory also includes Maxwell's theory with an interesting modification for curved space.

\vskip 4em
\large
{\bf \underline {Theory}}
\vskip 1em
As in GR we accept Einstein's "covariant" postulate. In other words, we assume that the description of 
matter should not depend on a system of coordinates. In mathematical terms it means that the theory
must be expressed in terms of tensor analysis. We also postulate that matter can be described by a
3-index tensor $P^{\, i}_{jk}$ without any additional assumptions on the symmetry of $P^{\, i}_{jk}$ with respect to the
lower indices. The tensor $P^{\, i}_{jk}$ serves as a potential for the tensor of matter.
In order to describe the change of the tensor-potential with respect to coordinates, we must introduce a covariant
derivative - and for that we need the metric tensor $g_{ij}$.
Since $P^{\, i}_{jk}$ are the only variables available to us, the metric $g_{ij}$ then should be defined by means of $P^{\, i}_{jk}$ only. 
The only way we can construct $g_{ij}$ out of $P^{\, i}_{jk}$ is to postulate that the metric of space, $g_{ij}$, can be expressed in terms
of the tensor potential by a quadratic algebraic relationship. The most general quadratic equation has this form:

\begin{eqnarray}
\label{f25}
g_{ij} = b_1{\bar P}^{\, m}_{ni}{\bar P}^{\, n}_{mj}+b_2{\hat P}^{\, m}_{in}{\hat P}^{\, n}_{jm}+
b_3({\bar P}^{\, m}_{in}{\hat P}^{\, n}_{mj}+{\bar P}^{\, m}_{jn}{\hat P}^{\, n}_{mi})+b_4{\bar P}_{\, m}{\bar P}^{\, m}_{ij} \nonumber\\
+\, b_5{\hat P}_m{\hat P}^{\, m}_{ij}+b_6{\bar P}_i{\bar P}_j+b_7{\hat P}_i{\hat P}_j+b_8({\bar P}_i{\hat P}_j+{\bar P}_j{\hat P}_i)
\end{eqnarray}

In the above equation, ${\bar P}$ and ${\hat P}$ are the symmetrical
and anti-symmetrical (low indices) parts of the tensor $P^{\, i}_{jk}$, respectively:

\begin{eqnarray}
\label{f26}
&&{\bar P}^{\, i}_{jk}=(P^{\, i}_{jk}+P^{\, i}_{kj})/2,{\quad}and{\quad}{\hat P}^{\, i}_{jk}=(P^{\, i}_{jk}-P^{\, i}_{kj})/2\\
&&{\bar P}_i={\bar P}^{\, k}_{ki},{\quad}and{\quad} {\hat P}_i={\hat P}^{\, k}_{ki}. 
\end{eqnarray}

We must require that the metric tensor $g_{ij}$ has its inverse $g^{ij}$.  Directly from this and from (\ref{f25}) we have these
important results: 
a) the metric is a function of matter ($P^i_{jk}$); the metric cannot exist without matter; b) nowhere can the tensor potential $P^{\, i}_{jk}$
be equal to zero - matter cannot disappear entirely; c) if $P^{\, i}_{jk}$ can be considered constant in some region, 
its associated space is flat ($g_{ij}=constant$).  From the latter it follows that the reason we live in an almost flat space
is because the gravitational matter ($P^{\, i}_{jk}$) associated with a distant massive object (e.g. galaxy), can be considered constant 
within the solar system.

Another attractive feature of this approach is that the definition of the metric tensor $g_{ij}$ allows a natural
approach for a perturbation theory. A small deviation on matter ($P^{\, i}_{jk}$) leads to a small deviation in the space metrics.

We will address the issue of defining the constants $b_1, ...,b_8$ in the sections below when we discuss particular matters
such as gravitation and electromagnitism. 
However, regarding the constants, one point has to be mentioned at this time: once they are chosen, the metric is defined in its absolute value.
In Eistein's GR, the metric tensor $g_{ij}$ is only defined up to a constant factor. This factor is chosen under the condition that at
infinity the metric $g_{ij}$ is the Minkowski metric: $g_{ij}=(1, -1,-1,-1)$. Unlike in GR, in this theory, the "bigger" the matter
($P^{\, i}_{jk}$), the "bigger" the metric tensor $g_{ij}$.

Both $P^{\, i}_{jk}$ and $g_{ij}$
are defined in a 4-pseudoRiemann space with $g_{ij}$ having a Minkowski signature (+ - - -). 
The tensor of matter $M^{\, i}_{jkl}$ can then be 
defined in the same spirit as the Riemann tensor with respect to the $\Gamma$s. In other words, the tensor of 
matter will have covariant derivatives of $P^{\, i}_{jk}$ (or in short $P^{\, \prime}$) and terms of the square of 
$P^{\, i}_{jk}$ (or in short $P^{\, 2})$:
\begin{equation}
\label{f2}
M=\Sigma (cP^{\, \prime} + {\bar c}P^{\, 2})
\end{equation}
Here, $\Sigma$ stands for the sum of terms with respect to some index permutations and $c$, ${\bar c}$ are sets of dimensionless constants.
This form is dictated by the scaling factors of  $P^{\, i}_{jk}$ and $M^i_{jkl}$: $P^{\, i}_{jk}$ is ~$cm^{-1}$ and $M^{\, i}_{jkl}$ ~$cm^{-2}$.
Examples of such a tensor M could be:

\begin{eqnarray}
\label{f2a}
&&a)M^i_{jkl}= P^{\, i}_{jk;l} \nonumber\\
&&b)M^i_{jkl}= 3P^{\, i}_{jk;l}+P^{\, i}_{mk}P^{\, m}_{jl} \\
&&c)M^i_{jkl}= P^{\, i}_{jl;k}-P^{\, i}_{jk;l}+P^{\, i}_{mk}P^{\, m}_{jl}-P^{\, i}_{ml}P^{\, m}_{jk} \nonumber
\end{eqnarray}
The last expression is similar in form to the Riemann tensor with respect to $\Gamma$s.
It is important to note that the tensor M does not include the  geometric Riemann tensor $R^{\, i}_{jkl}(g_{ij})$,
which only depends on the metric tensor $g_{ij}$. This can be justified by the fact that the Lagrangian should depend
only on first derivatives of independent variables. The Riemann tensor ($R^{\, i}_{jkl}$), on the other hand,
depends on second derivatives of $g_{ij}$ and thus on second derivatives of $P^{\, i}_{jk}$.

The equation for the distribution of matter is obtained from the variational principle. The Lagrangian here depends
on tensors $M^{\, i}_{jkl}$ and $g_{ij}$, which both are functions of the tensor potential $P^{\, i}_{jk}$ only.
\begin{equation}
A=\int\nolimits L(M^{\, i}_{jkl}, \, g_{ij}){\,} {\sqrt g} d^4x
\label{f3}
\end{equation}
where $g=-det(g_{ij})$.

By analogy with other flat space field theories, we will assume that the Lagrangian is quadratic with respect to the matter tensor M.
Because the Lagrangian L is a function of $P^{\, i}_{jk}$ through the square of the matter tensor M, it can include only three types
of terms: a) the square of the derivative of $P^{\, \prime} (L={P^{\, \prime}}^{\, 2})$, b) the derivative of P $(P^{\, \prime})$ and the square of P 
$(L=P^{\, \prime}P^{\, 2})$ and c) the 4th power of P, $(L= P^{\, 4})$. We shall reasonably assume that the reverse is also true: 
if we have a  Lagrangian that has only terms of type a), b), and c), we can always create a tensor of matter (${M^{\, i}}_{jkl}$)
in such a way that the Lagrangian can be written as a square function of tensor M (and $g_{ij}$, of course).
Thus, analyzing different forms of the Lagrangian, we will only be concerned with the fact that its terms satisfy 
one of the three above mentioned requirements.

We can introduce Lagrange coefficients (or Lagrange multipliers) $T^{\, ij}$,
and consider $g_{ij}$, $P^{\, i}_{jk}$ and $T^{\, ij}$ as independent variables. In this case the action integral A can be written as:

\begin{equation}
\label{f28}
A=\int\nolimits \sqrt{g}{\,} d^4x{\,} \{L(P^{\, i}_{jk;l}, P^{\, i}_{jk}, g_{ij})+T^{\, ij}[g_{ij}-Q_{ij}(P^{\, i}_{jk})]\}
\end{equation}

Here, $Q_{ij}(P^{\, i}_{jk})$ is a quadratic function given by eq.(\ref{f25}).
The variational principle will yield these three sets of equations
(Appendix A):
\begin{equation}
\label{f29}
\frac {{\delta}A} {{\delta}P^{\, i}_{jk}}=0{\quad}or {\quad}
{\partial}L/{{\partial}P^{\, i}_{jk}}-({\partial}L/{{\partial}P^{\, i}_{jk;l})}_{;{\,}l}-T^{\, mn}[{\partial}Q_{mn}/{\partial}P^{\, i}_{jk}]=0
\end{equation}

\begin{equation}
\label{f29a}
\frac {{\delta}A} {{\delta}g_{ij}}=0{\quad}or{\quad}  {\frac {{\partial}L} {{\partial}g_{ij}}}+{\frac{1}{2}}{\,} Lg^{ij}+{\frac{1}{4}}{\,}
{[-J^{\, ijs}-J^{jis}+J^{sij}]}_{;s}+T^{\, ij}=0
\end{equation}
In equation (\ref{f29a}), we indroduced the tensor $J^{\, ijs}$ which is defined by the following expression:
\begin{eqnarray}
\label{f29b}
&&J^{ijs}=g^{sr}[({L_r}^{mni}P^j_{mn}-{L_m}^{jni}P^m_{rn}-{L_m}^{nji}P^m_{rn})\nonumber\\
&&+({L_s}^{mnj}P^i_{mn}-{L_m}^{inj}P^m_{sn}-{L_m}^{nij}P^m_{sn})] \nonumber\\
&&{L_i}^{jkl}={\partial}L/{\partial}P^i_{jk;l}
\end{eqnarray}

\begin{equation}
\label{f29c}
\frac { {\delta}A}{{\delta}T^{ij}}=0 {\quad} or {\quad} g_{ij}-Q_{ij}(P^i_{jk})=0 \\
\end{equation}
Obviously, the set of Lagrange multipliers $T^{ij}$, defined by eq. (\ref{f29a}), is the energy-momentum tensor of matter.
It is interesting to know if the four Noether identities derived from the 2nd Noether theorem would lead to the expression ${T^{ij}}_{;j}=0$.

Let us now discuss a situation of "small" matter in a field of "large" matter. An example here could
be an electron in the field of Earth or even our own galaxy. The tensor of the potential P now can be written as a
superposition of  $P_o$ and p of "large" matter and "small" matter, respectively: $P=P_o+p{\,}(p \ll P_o)$.  
The metric tensor $g_{ij}$, based on eq. (\ref{f25}), can be written as $g_{ij}=g_{oij}+{\bar g}_{ij}(P_o,p), \,\,(\, {\bar g}_{ij}\ll g_{oij})$.
Because $P_o$ and $g_o$ are fields of the "large" matter, their 
change on the scale of the "small" matter can be ignored. In this case, the background potential $P_o$ can be
considered a constant and the space $g_o$ is flat. 

We now can apply a perturbation method to
the Lagrangian leaving only linear and quadratic terms with respect to $p^i_{jk}$ and ${\bar g}_{ij}$.
The linear terms are responsible for the 
decay of the "small" matter through interaction with the "large" matter. The quadratic terms
describe the state of "small" matter in the flat space. These terms include
constants ${P_o}^i_{jk}$ and the Minkowski metric (1, -1, -1, -1). ${P_o}^i_{jk}$ now serves as a set of parameters of the "small" matter
(e.g. mass, charge, etc.). Thus, the values of the background field of the "large" matter determine the parameters of the "small" matter. 
In other words, the "large" matter gives the basis and scale for the values of the "small" matter.
The matter tensor M expressed by equation (\ref{f2}) in the case of the "small" matter perturbation can be written in this symbolic form: 
$M=M_o+p{\, \prime} +P_op$. What follows from this is the realization that the characteristic length of the "small" matter p is $\approx$ $1/P_o$.
If we think that the characteristic length of an elementary particle (e.g. proton) is about $10^{-13}cm$,
then $P_o$ is about $10^{13} {\,}cm^{-1}$.
The stronger the background field is (larger $P_o$), the smaller the "elemetary" particle.

The linearized theory of "small" matter can also be viewed as a transition to quantum field theory.
It is not obvious if the linearized quadratic Lagrangian is an accurate description of the P-potential
of elementary particles at a distance much less than $1/P_o$. This inaccuracy might present mathematical problems similar to those 
of contemporary quantum mechanics.
\vskip 2em

{\bf \underline {Law of Conservation of Matter}}
\vskip 1em
In the covariant description of physics, the law of conservation must be written in a form where the divergence
of a vector is zero ($J^k_{;k}=0$). This leads directly to a mathematical fact that the total flux of a vector 
$J_k$ through any closed hyper-surface of 4D space is zero \cite{i10}, \cite{i12},\cite{i13}.
We will now show that the equations of motion (or equations for $P^i_{jk}$) contain the covariant law of conservation,
which we will call the law of conservation of matter. This law comes directly from the equation (\ref{f29a}) by contracting its indices.

\begin{eqnarray}
\label{f5a}
&&g_{ij}\frac {{\delta}A}{{\delta}g_{ij}}=0{\quad} or \nonumber\\
&&g_{ij}[{\frac {{\partial}L} {{\partial}g_{ij}}}+{\frac{1}{2}}{\,} Lg^{ij}]+{\frac{1}{4}}{\,}
g^{ij}{[-J^{ijs}-J^{jis}+J^{sij}]}_{;s}+T=0,\\
&& where   \nonumber\\
&&J^{ijs}=g^{sr}[({L_r}^{mni}P^j_{mn}-{L_m}^{jni}P^m_{rn}-{L_m}^{nji}P^m_{rn}) \nonumber\\
&&+({L_s}^{mnj}P^i_{mn}-{L_m}^{inj}P^m_{sn}-{L_m}^{nij}P^m_{sn})] \nonumber\\
&& {L_i}^{jkl}={\partial}L/{\partial}P^i_{jk;l}  \nonumber
\end{eqnarray}

The expression in the first square bracket of eq. (\ref{f5a}) vanishes due to the fact that L is quadratic with respect to $M^{\,i}_{jkl}$ (or $g^{ij}$).
The remaining part of this equation can be written in the form:

\begin{eqnarray}
\label{f5b}
&&{J^s}_{;s}+T=0 \nonumber\\
&&J^s=g_{ij}g^{sr}[({L_r}^{mni}P^j_{mn}-{L_m}^{jni}P^m_{rn}-{L_m}^{nji}P^m_{rn})+\nonumber\\
&&{\quad}({L_s}^{mnj}P^i_{mn}-{L_m}^{inj}P^m_{sn}-{L_m}^{nij}P^m_{sn})] \nonumber\\
&& {L_i}^{jkl}={\partial}L/{\partial}P^i_{jk;l}
\end{eqnarray}

If $T=0$, the expression (\ref{f5b}) represents the law of conservation. Thus, we will postulate that for any physically correct Lagrangian
this must be the case.

Due to the fact that any Lagrangian has a gauge of a coordinate transformation, which leads to the four Noether identities, 
it is possible that the equations (\ref{f29} thru \ref{f29c}) do not form a complete set. If this is the case, the equation $T=0$ can be added 
as an additional physical postulate. Another way to assure that $T=0$ is to consider a Lagrangian that satisfies a certain U(1) gauge.
This reduces the number of equations by one, thus allowing the addition of one more equation of $T=0$ to complete the set.

\vskip 5em

{\bf \underline{Physical Identification of $P^i_{jk}$}}
\vskip 1em

We can now address the question of what physical meaning should be assigned to the tensor $P^i_{jk}$ (e.g. gravitation, electromagnetism, etc.).
We will make this identification based on the symmetry of the tensor potential $P_{ijk}$, 
obtained from $P^i_{jk}$ by lowering its first index.
If we accept the assumption that $P_{ijk}$ has no preset symmetry, then $P_{ijk}$
can be written in the following form (Appendix B):
\begin{eqnarray}
\label{f8}
&&P_{ijk}=[G_{ijk}+1/6(G_ig_{jk}+G_kg_{ij}+G_jg_{ki})] + \nonumber\\
&&+[(B_{jki}+B_{kji})+1/3(B_jg_{ik}+B_kg_{ij}-2B_ig_{jk})]\nonumber\\
&&+ [C_{ijk}+1/3(C_kg_{ij}-C_jg_{ik}) ]+ 1/6{\epsilon}_{ijkm}E_m 
\end{eqnarray}
In the above expression, $\epsilon_{ijkl}$ is a fully anti-symmetrical tensor of Levi-Civita.

Here, we have introduced four new vectors ($G_i$, $B_i$, $C_i$, $E_i$) and 
three new tensors ($G_{ijk}$, $B_{ijk}$, $C_{ijk}$).
The tensor $G_{ijk}$ is fully symmetrical, and its trace is equal to zero $g^{km}G_{ikm}=0$; thus, it has 16 independent components. 
$B_{ijk}$ is anti-symmetrical in the 2nd and 3rd indices and it is trace free: $g^{ij}B_{ijk}=0$.
This tensor satisfies one more condition: $\epsilon^{ijkm}B_{ijk}=0$ (Appendix B).
From here it follows that $B_{ijk}$ also has 16 independent components.
$C_{ijk}$ is also anti-symmetrical in the 2nd and 3rd indices and has the same property as $B_{ijk}$. 
Namely, $g^{ij}C_{ijk}$=0, $\epsilon^{ijkm}C_{ijk}=0$ and thus it has 16 independent components as well.

The representation given by expression (\ref{f8}) is reversible. We can express every vector $G_i$, $B_i$, $C_i$, and $E_i$
as well as tensors $G_{ijk}$, $B_{ijk}$ and $C_{ijk}$ through the original tensor potential $P_{ijk}$ and the metric tensor
$g_{ij}$ (Appendix B).

Due to its reversibility, we can formulate the theory in terms of these eight new sets of variables 
($G_{ijk}$, $B_{ijk}$, $C_{ijk}$, $G_i$, $B_i$, $C_i$, $E_i$, $g_{ij}$) with some care paid to the derivation of eq. (\ref{f29a})
to be sure that the law of conservation of matter is maintained. Five more Lagrange coefficients (or Lagrange multipliers)
must be added to the Lagrangian to reflect the traceless properties of $G_{ijk}$, $B_{ijk}$, $C_{ijk}$ tensors.

\vskip 5em

{\bf \underline{Gravitation}}
\vskip 1em

The simplest element in expression (\ref{f8}) is a vector and we will identify one of the four vectors with gravitational matter.
It is not difficult to show that matter described by the vector $E_i$ (or fully anti-symmetrical $P_{ijk}$) cannot form space. 
If $P^i_{jk}=1/6{\epsilon}^i_{jkm}E^m$, then the expression for $g_{ij}$ (\ref{f25}) has only one term:
\begin{eqnarray}
\label{f38}
&&g_{ij}=b_2/36{\hat P}^m_{nik}{\hat P}^n_{mjk}=b_2/36{\epsilon}^m_{njk}{\epsilon}^n_{mil}E^kE^l \nonumber\\
&&=b_2/18[g_{ij}(E^kE_k)-E_iE_j].
\end{eqnarray}
It follows from eq. (\ref{f38}) that $g_{ij}$, at any point, has a degenerate form with its determinant equal to zero: $det(g_{ij})=0$.

Similarly, vector $C_i$ cannot form space by itself. In this case, matter is described only by vector $C_i$ and metric $g_{ij}$:
\begin{equation}
\label{f39}
P^i_{jk}={\frac{1}{3}}(C_kg_{jm}-C_jg_{mk})g^{im}
\end{equation}
The expression for $g_{ij}$ then takes this form:
\begin{equation}
\label{f40}
g_{ij}=b_2{\hat P}^m_{ni}{\hat P}^n_{mj}+b_7{\hat P}_i{\hat P}_j=(b_2/3+b_7)C_iC_j
\end{equation}
This obviously does not satisfy the condition: $det(g_{ij}){\neq}0$.
This leaves us with only two possibilities, $G_i$ and $B_i$. The vector $G_i$ appears to be a more appropriate candidate.
As we will see later, by choosing constants $b_1$, $b_4$ and $b_6$ in such a way that field $G_i$ can form space, we simultaneously
forbid vector $B_i$ to exist by itself.

If we choose $G_i$ as a vector representing gravitational matter, the tensor potential of the pure gravitational field
should only be expressed through vector $G_i$ and metric $g_{ij}$. 
\begin{equation}
\label{f40a}
P_{ijk}=(G_ig_{jk}+G_jg_{ik}+G_kg_{ij})/6
\end{equation}
The expression for $g_{ij}$, eq. (\ref{f25}), takes the form of:
\begin{equation}
\label{f40b}
g_{ij}=b_1{\bar P}^m_{ni}{\bar P}^n_{mj}+b_4{\bar P}_m{\bar P}^m_{ij}+b_6{\bar P}_i{\bar P}_j
\end{equation}
Substituting (\ref{f40a}) in (\ref{f40b}) we get:
\begin{equation}
\label{f40c}
g_{ij}=(2b_1/36+b_4/6)(G_kG^k)g_{ij}+1/36(10b_1+12b_4+36b_6)G_iG_j
\end{equation}
In order for $g_{ij}$ to be non-degenerate, the last term containing $G_iG_j$ must vanish, thus requiring that
$b_1$, $b_4$ and $b_6$ satisfy
the condition:
\begin{equation}
\label{f40d}
(10b_1+12b_4+36b_6)=0
\end{equation}
This is only one equation for the three unknown $b_1$, $b_4$ and $b_6$.
However, there are other considerations that put additional restrictions on the choice of $b_1$, $b_4$ and $b_6$ coefficients.
These restrictions accur because the above representation of the fully symmetrical tensor $P_{ijk}$ through vector $G_i$ is not complete.
The most general expression can be written as:
\begin{equation}
\label{f41}
P_{ijk}=g_{im}P^m_{jk}=(G_ig_{jk}+G_jg_{ik}+G_kg_{ij}+k_0G_iG_jG_k)/(6+k_1)
\end{equation}
where $k_1=k_0G^kG_k$.

Substituting (\ref{f41}) in (\ref{f40b}), we get this expression for $g_{ij}$:
\begin{eqnarray}
\label{f43}
&&g_{ij}=b_1/{(6+k_1)}^2\,[(G_ig_{mn}+G_mg_{in}+G_ng_{im}+k_0G_iG_mG_n) \nonumber\\
&&\quad \quad \quad (G_jg_{kl}+G_kg_{il}+G_lg_{ik}+k_0G_iG_kG_l)g^{mk}g^{nl}]+ \nonumber\\
&&b_4/(6+k_1)\,G_n(G_ig_{jm}+G_jg_{im}+G_mg_{ij}+k_0G_iG_jG_m)g^{mn}+b_6G_iG_j\nonumber\\
&&=[2b_1/{(6+k_1)}^2+b_4/(6+k_1)]G^mG_mg_{ij}+\nonumber\\
&&[{k_1}^2(b_1+b_4+b_6)+k_1(6b_1+8b_4+12b_6)+(10b_1+12b_4+36a_6)]G_iG_j\nonumber\\
\end{eqnarray}
In order for $g_{ij}$ to satisfy this equation and to satisfy the requirement $det(g)\neq0$, the term in the square bracket
in front of $G_iG_j$ should be zero and the factor in front of $g_{ij}$ on the right hand side should be one.

\begin{eqnarray}
\label{f44}
&&[2b_1/{(6+k_1)}^2\, g_{ij}+b_4/(6+k_1)]G^mG_m=1\\
\label{f44a}
&&{k_1}^2(b_1+b_4+b_6)+k_1(6b_1+8b_4+12b_6)\nonumber\\
&&+(10b_1+12b_4+36b_6)=0
\end{eqnarray}

Eq. (\ref{f44a}) is a quadratic equation with respect to $k_1$.
Thus, for any given $b_1, b_4, b_6$ we can find two independent values for $k_1$ (and subsequently $k_0$), which would correspond
to two independent gravitational fields. The possibility of having two gravitational fields is obviously unacceptable.
Thus, $b_1$, $b_4$, $b_6$ should be such that they assure a single solution in which $k_1=0$. 
From eq. (\ref{f44}), it follows that $G^mG_m$ is a constant. We can always require that this constant is equal to one: $G^mG_m=1$.

There are two sets of $b_1$, $b_4$, $b_6$ that satisfy the above requirements: a) $b_1=-18$, $b_4=12$, $b_6=1$ and b) 
$b_1=-8$, $b_4=26/3$, $b_6=-2/3$. The first solution is more acceptable from an aesthetic point of view.

It is interesting  to note that in the case of ($b_1$, $b_4$, $b_6)=(-18, 12, 1)$, it can be shown that if matter (tensor potential $P_{ijk}$)
consists only of the gravitational tensor $G_i$ and the fully symmetrical tensor $G_{ijk}$, eq. (\ref{f25}) demands that $G_{ijk}=0$.

If we are to look for a  Lagrangian L that describes a pure gravitational field, we have to consider it to be a function only
of the gravitational field $G_i$ and the metric $g_{ij}$. 
\begin{eqnarray}
\label{f45}
&&A=\int\nolimits dx^4 {\sqrt g}{\,} [k_1G^{\, k;l}G_{k;l}+k_2G^{\, l;k}G_{k;l}+k_3({{G^{\, k}}_{;k}})^2 \nonumber\\
&&+k_4{G^{\, k}}_{;k}G^{\, l}G_l+T(G^{\, k}G_k-1)]
\end{eqnarray}
Here we limit the Lagrangian to only four types of terms as described above in eq. (\ref{f2a}a, b, c). 
We omitted the term ${(G^kG_k)}^4$ to assure that $G_i=constant$ is a solution.

In the last ten years, a significant number of papers were written where a similar Lagrangian was considered (\cite{i15}, \cite{i16}).
The focus of these papers was the "violation of the Lorentz invariance", which was achieved by adding, in addition to a space metric,
a vector field representing gravitational matter and its constraint ($G^kG_k=1$). If the addition of a vector field as gravitational matter
is prompted from physical considerations, the addition of the constraint ($G^kG_k=1$), however, is  done without any physical
or mathematical justification.

In contrast, in our theory, the gravitational vector $G_i$ and its constraint ($G^kG_k=1$) are consequences of the simplified form of $P^i_{jk}$
when it represents only the gravitational field ($G_i,\,g_{ij}$).

With three independent variables (the gravitational field $G_i$, the metric $g_{ij}$ and the Lagrange multiplier T),
the set of equations can be written in the following form:
\begin{eqnarray}
\label{f55a}
&&{\partial}A/{\partial}G_i=0 {\,}{\,}or\nonumber\\
&&-k_1{G_{i;k;}}^k-k_2{G_{k;i;}}^k-k_3{{G_{k;}}^k}_{;i}+k_4{{G_{k;}}^k}G_i+TG_i=0
\end{eqnarray}

\begin{eqnarray}
\label{f55b}
&&{\partial}A/{\partial}g_{ij}=0 {\,}{\,}or{\,}\nonumber\\
&&k_1[-{G^k}_{;i}G_{k;j}-G_{i;k}{G_{j;}}^k+1/2g_{ij}{G^m}_{;n}{G_{m;}}^n]+\nonumber\\
&&k_2[-{G^k}_{;i}G_{j;k}-{G^k}_{;j}G_{i;k}+1/2g_{ij}{G^m}_{;n}{G^n}_{;m}]+\nonumber\\
&&k_3[-{G^k}_{;k}(G_{i;j}+G_{j;i})+1/2g_{ij}{({G^k}_{;k})}^2]+\nonumber\\
&&k_4[-G_iG_j{G^{\, k}}_{;k}+1/2g_{ij}{G^{\, k}}_{;k}]+\nonumber\\
&&{\frac {(k_1+k_2)}{2}} {{[(G_{i;k}+G_{k;i})G_j+(G_{j;k}+G_{k;j})G_i-(G_{i;j}+G_{j;i})G_k]}_;}^k+\nonumber\\
&&k_3[{({G^k}_{;k}G_i)}_{;j}+{({G^k}_{;k}G_j)}_{;i}-g_{ij}{({G^k}_{;k}G^m)}_{;m}]+\nonumber\\
&&+k_4/2[G_{i;j}+G_{j;i}-g_{ij}{G^{\, k}}_{;k}]-TG_iG_j=0
\end{eqnarray}

\begin{equation}
\label{f55c}
{\partial}A/{\partial}T=0 {\,}{\,}or{\,}{\,} G^kG_k=1
\end{equation}

An additional condition is taken as a postulate:
\begin{equation}
\label{f55d}
T=0
\end{equation}

Can a Lagrangian be identified (constants $k_1$, $k_2$, $k_3$ and $k_4$) in such a way that the solution for $g_{ij}$
is equal to the Schwarzschild solution, or at least so at large distances (first order of 1/r)?

Let us consider a particular case, when $k_1=0,\,\,k_2=1,\,\,k_3=-1$ and $k_4=0$. In this case the Lagrangian has this form:

\begin{equation}
\label{f9}
L=(G_{i;j}G_{l;k}-G_{i;k}G_{j;l})g^{ik}g^{jl}+T(G^kG_k-1)
\end{equation}

A variation of action A ($A=\int\nolimits L(G_i, g_{ij}){\,} {\sqrt{g}} d^4x$) by $G_i$ leads to this equation:
\begin{equation}
\label{f10}
-g^{km}(G_{k;i;m}-G_{k;m;i})+TG_k=-R_{im}G^m+TG_i=-R_{im}G_m=0
\end{equation}
Here, $R_{im}$ is the Ricci tensor and we have used the postulate T=0.
The second set of equations is obtained through a variation by $g_{ij}$ and has the following expression
(Appendix C):
\begin{eqnarray}
\label{f11}
&&-G_{i;m}{G^m}_{;j}-G_{j;m}{G^m}_{;i}+{\frac{g_{ij}}{2}}{ }G_{k;m}G^{m;k}+ \nonumber\\
&&+(G_{i;j}+G_{j;i}){G^m}_{;m}-{\frac{g_{ij}}{2}} {({G^m}_{;m})}^2+          \nonumber\\
&&+{\frac{1}{2}}{[(G_{i;s}+G_{s;i})G_j+(G_{j;s}+G_{s;j})G_i-(G_{i;j}+G_{j;i})G_s]_;}^s- \nonumber\\
&&-{G^m}_{;m}(G_{i;j}+G_{j;i})+{({G^m}_{;m}G_s)_;}^sg_{ij}-Tg_{ij}=0
\end{eqnarray}

Contracting eq. (\ref{f11}), and remembering that T=0, we will get the law of conservation of gravitational matter:
\begin{equation}
\label{f12}
{J_{k;}}^k=0, \quad where \quad J_k=G_{m;k}G^m+G_{k;m}G^m+G_k{G^m}_{;m}
\end{equation}

Since all equations for $G_i$ and $g_{ij}$ include only covariant derivatives, the constant $G_i$ and the constant $g_{ij}$
are the solutions  of eqs. (\ref{f10}, \ref{f11}).
Metric $g_{ij}$, by choice of a system of coordinates, can be made to have a Minkowski form, $g_{ij}=(1,-1,-1,-1)$, at any point.
Vector $G_i$ should be a time vector due to the constraint $G_kG^k=1$.

Let us consider a stationary (time independent) spherical symmetrical solution that at infinity becomes a constant,
as described in the paragraph above. 

In this case, the vector $G_i$ has only two components: $G_0$ ($G_t$) and $G_1$ ($G_r$), which are both 
functions of only the distance "r". From textbooks it is well known that $R_{ij}$
(Ricci tensor) has three non-zero components: $R_{00}$ ($R_{tt}$), $R_{11}$ ($R_{rr}$) and $R_{22}$ ($R_{{\theta}{\theta}}$)
=$R_{33}$($R_{{\phi}{\phi}}$) (\cite{i10},\cite{i12},\cite{i13}).

The vector equation (\ref{f10}) ($R_{im}G^m=0$) then can be written as two independent equations:
\begin{equation}
\label{f12a}
R_{00}G_0=0 \quad and \quad R_{11}G_1=0
\end{equation}

From this it follows that $R_{00}=R_{11}=0$. For a stationary
spherical solution it can be shown that these two equations are equivalent to $R_{ij}=0$ (\cite{i12}, p. 274). Thus, the $g_{ij}$ in our
solution is the Schwarzschild metric.

In order to find the functions $G_0$ and $G_1$, we will use two equations. The first one is the equation of constraint:
\begin{equation}
\label{f12b}
G_kG^k=1\,\, or \,\, G_0g^{00}+G_1g^{11}=1
\end{equation}
The second equation is the equation of the conservation of matter.

\begin{eqnarray}
\label{f12c}
&J^k_{;k}=0 \quad or \quad ({J_1g^{11}\sqrt g)}_{,1}=0
\end{eqnarray}
From these two equations, we can obtain:
\begin{eqnarray}
\label{f12d}
&&G_1=C_1/(r-r_o)\,\,and\,\, G_0=\sqrt{1-r_o/r+{C_1}^2/r^2}\nonumber\\
&&g_{00}=(1-r_o/r),\,\,g_{11}=-1/g_{00}
\end{eqnarray}
In the expression above, $C_1$ is a constant.

Let us notice, that from an experimental point of view, the two main verifications of GR - the deflection of light by the Sun and
the shift of the perihelion of Mercury - are all derived from Einstein's vacuum equation. To be more precise,
they are both derived from the Schwarzschild metric, $dG^2=(1-r_0/r)dt^2-1/(1-r_0/r)dr^2-R^2d\Omega^2$.  
To satisfy both of these experimental facts, when matter is reduced to only gravitational matter ($G_i$, $g_{ij}$ ),
our theory has to yield a Schwarzschild metric $g_{ij}$ as the solution. As we have shown above (eq. \ref{f12a}), this is the case.

Just as in GR, we postulate that a "small" macroscopic solid body, as well as a ray of light, moves along geodesic lines of
$g_{ij}$.

In the case of an arbitrary gravitational field $G_i(x^i)$, $g_{ij}(x^i)$, it is important to note that the equations
(\ref{f10}, \ref{f11}) do not look like Einstein's equations $R_{ij}=kT_{ij}$.  However, Einstein's vacuum equation, $R_{ij}=0$, is 
sufficient to satisfy the first one (\ref{f10}). We can not reject the the possiblity that $R_{ij}=0$ is the only metric solution 
for eqs. (\ref{f10}, \ref{f11}) for any "vacuum" problem (i.e. gravitational matter ($G_i$) and metric $g_{ij}$ only). 

When the constant $C_1$ approaches zero, we will have a simplified expression for vector field $G_k$: $G_0=\sqrt{g_{00}}$ and $G_1=0$.
As we can see from this solution, the closer we get to the mass, the lower the gravitational matter becomes. In other words, the gravitational
matter is replaced by "hard" matter. That also means that a gravitational field must exist independently of matter. 
So it appears that we are back to the ether theory. The ether in this case is the gravitational matter 
described by a vector field. The entire Universe if filled by the gravitational matter. Elementary particles are the depressions
in gravitational matter or might even be true singularities.
 
We have to investigate the other vector Lagragians to see if another chioce of constants, $k_1, k_2, k_3, k_4$ in eq. (\ref{f45}),
can lead to the same Schwarzschild solutions (at least at large distances), but do not lead to the ether theory. 
Such solutions should have $G_0$ that increases towards the center of the mass and be able to go to zero at infinity.

\vskip 1em
\vskip 1em
{\bf \underline {Electromagnetism}}
\vskip 1em
We now address the question of electromagnetism, which will be identified with the vector field $E_i$.
As we showed earlier, the electromagnetic (EM) field by itself cannot form space.
Let us consider a situation when we have gravitational and EM fields.
Then, $P^i_{jk}$ has this form:
\begin{eqnarray}
\label{f47a}
&&P^i_{jk}=g^{im}P_{mjk}\quad and {\quad} \nonumber\\
&&P_{ijk}=1/6(G_ig_{jk}+G_jg_{ik}+G_kg_{ij}+{\epsilon}_{ijkm}E^m)
\end{eqnarray}
Substituting expression (\ref{f47a}) into (\ref{f25}) and  taking ($b_1$, $b_4$, $b_6$) as (-18, 12, 1) we get the following relation:
\begin{equation}
\label{f48a}
g_{ij}=(G^kG_k)g_{ij}+b_2(g_{ij}-E_iE_j)
\end{equation}
Eq. (\ref{f48a}) does not satisfy the requirement $g=-det(g_{ij})\neq 0$ unless $b_2=0$.

Though the EM field is not explicitly present in eq. (\ref{f25}), it should not be taken to mean that the space metric does not depend
on the EM field. The dependence of the space metric on the EM field exists through the equations of motion which are
derived from the Lagrangian.

In the case of only gravitational $G_i$ and EM field $E_i$, the constraint on the magnitude of 
the gravitational field remains the same: $G^kG_k=1$. In this case, both the metric $g_{jk}$ and the gravitation vector $G_i$ 
are influenced by the presence of the EM field in accordance with the equations of motion.

Since the constant $b_2$ in equation (\ref{f48a}) is zero, it can be assumed that the metric tensor $g_{ij}$
depends only on the symmetrical, in low indices, components of the tensor potential $P^i_{jk}$.
In this case, the metric $g_{ij}$ as a function of $P^i_{jk}$ is fully defined by the expression:
\begin{equation}
\label{f48}
g_{ij} = -18{\bar P}^m_{ni}{\bar P}^n_{mj}+12{\bar P}_m{\bar P}^m_{ij}+{\bar P}_i{\bar P}_j
\end{equation}
Here ${\bar P}$ is the symmetrical part of the tensor $P^i_{jk}:\,\,{\bar P}^i_{jk}=(P^i_{jk}+P^i_{kj})/2$

The Lagrangian of the EM field should be similar to that of Maxwell's theory in flat space:
\begin{equation}
\label{f16}
L_{em}=E_{i;j}E_{m;n}g^{im}g^{jn}
\end{equation}
In eq. (\ref{f16}), $E_{ij}=E_{i;j}-E_{j;i}$.

As we mentioned above, for the EM field to exis,t it needs another field - a gravitational field.
Thus, writing the Lagrangian for the EM field, we MUST include the gravitational field as well.
The total Lagrangian then can be written as a sum of three Lagrangians: $L_{em}$ - Maxwell, $L_g$ - gravitational
(see section above) and $L_{g-em}$ - electromagnetic-gravitational interaction.

\begin{equation}
\label{f17}
L=L_{em}(E_{ij}{\,},{\,}g_{ij})+L_g(G_i{\,},{\,}g_{ij})+L_{g-em}(E_i{\,},{\,}G_i{\,},{\,}g_{ij})
\end{equation}
Based on expressions (\ref{f2}) and (\ref{f3}), $L_{g-em}$ could contain linear terms with respect to $E_{i;j}$
or quadratic terms with respect to $E_i$.
It is not difficult to show that the linear term (only one is possible) vanishes. Indeed, the action integral A with 
the linear term has this form:
\begin{eqnarray}
\label{f17a}
&&\int\nolimits d^4 \sqrt{g}\,G_{i;j}E_{k;l}{\epsilon}^{ijkl}=-\int\nolimits d^4 \sqrt{g}\,G_{i;j;l}E_k {\epsilon}^{ijkl}= \nonumber\\
&&-\int\nolimits d^4 \sqrt{g}\,R_{jlis}G^s{\epsilon}^{ijkl}E_k=0
\end{eqnarray}
Above, we have used the Bianchi identity for the Riemann tensor.
The quadratic terms with respect to $E_i$ amount to only four possibilities:
\begin{equation}
\label{f18}
L_{g-em}=k_1(G_{i;j}E^iE^j-{G_{k;}}^kE_jE^j)+k_2G^iG_iE^jE_j+k_3{(G^iE_i)}^2 
\end{equation}
$k_1$, $k_2$, and $k_3$  are dimensionless constants. The presence of the last two terms is equivalent to the statement that a photon
has a  very large mass proportional to the square of the external gravitational potential. This is a fact not supported by experimental data.
Therefore, we will ignore the terms with $k_2$ and $k_3$.

The total sets of equations can now be obtained
through variation of the Lagrangian with respect to the gravitational field $G_i$, electromagnetic field $E_i$ and metric $g_{ij}$.
The Maxwell equation will be modified by adding electromagnetic-gravitational interaction:
\begin{equation}
\label{f19}
\frac { {\partial}A}{{\partial}E_i}=0\quad or\quad {{E_i}^j}_{;j}=k_1[E^m(G_{i;m}+G_{m;i})-E^iG_{m;m}]
\end{equation}

Strictly speaking EM-gravitational interaction still exists even if we omit the $L_{g-em}$ term from expression (\ref{f17}).
This implicit interaction is through the space metric $g_{ij}$. The metric $g_{ij}$ and the gravitational matter $G_i$
must adjust themselves to satisfy the equations of motion. However, it seems fitting to accept the explicit
form of EM-gravitational interaction ($k_1 \neq 0$). In this case, a photon has a mass
that is proportional to the curvature of space or non-uniformity of the gravitational field $G_i$.
The important consequence of this interaction is that light emitted by a star will slowly
give up its energy to the gravitational field as it travels through space which manifests itself in the reduction of light frequency.
The longer the distance the light travels, 
the more frequency shift it will produce. That is the reason why we see far removed stars
(and galaxies) "moving" away from us. The farther they are, the "faster" (it appears) they move. 
This conclusion is solely based on the shift in the frequency of light due to the electromagnetic-gravitational interaction.

These considerations about electromagnetic-gravitational interactions strongly contradict the present theory of an "expanding" Universe
which is utilized for explaning the red-shift of distant stars.

The same argument also leads to another conclusion: light is not only deflected by a massive star (Sun), but it also should
change its frequency, when passing near it.

The dimensionless constant $k_1$ of this interaction can be estimated from the observation of light intensity
(or distance from the star) vs. frequency shift.

If we assume that the Universe' space metric (x-y-z) is a closed 3-D sphere of radius R, then light might travel several times 
across the Universe, thereby creating in effect a noise-like background radiation. 
This should allow us to estimate the size of the Universe.

\vskip 3em
{\bf \underline{The Cosmological Problem}}
\vskip 1em
In our theory, the cosmological problem becomes a rather simple one. 
The (x-y-z) space should be a 3-dimensional sphere in flat 4-D Euclidian space.
$x^2+y^2+z^2+w^2={R(t)}^2$, so that the metric can be expressed as:
\begin{equation}
g_{ij}=dt^2+R(t)^2({d\chi}^2+sin^2{\chi}d{\theta}^2+sin^2{\chi}sin^2{\theta}d{\phi}^2)
\end{equation}
The gravitational vector $G_i$ must have only a "t" (or $G_0$) component and the condition $G_kG^k=1$
leads to $G_0=1, G_{\chi}=G_{\theta}=G_{\phi}=0$.
In this case, it is not difficult to show that $R(t)=R_0=constant$ satisfies all the equations (\ref{f55a}, \ref{f55b}, \ref{f55c}).

Thus, the picture of the world is a static Universe filled with gravitational matter except for some areas
where the gravitational matter is replaced by other forms of matter (hard particles). Of course, one can ask the question:
why did the perfectly uniform gravitational matter become disturbed? And, what motivated this disturbance? 
Perhaps there is more than one Universe that can actually touch or even collide with each other causing nonuniformity or creating hard matter.
One might think of it as a water drop in weightless space colliding with another such drop.

\vskip 3em
{\bf \underline{Other Forms of Matter}}
\vskip 1em

Above we discussed two vector fields that could be identified with gravitational and electromagnetic matters. 
So what is the meaning of the other fields, $B_i$ and $C_i$? And what is the meaning of tensor fields $G_{ijk}$, $E_{ijk}$ and $T_{ijk}$? 
It seems that the latter ones should be identified with hard matter, such as neutrons, protons and electrons. 
It is tempting to identify these fields directly with elementary particles, or with their class - such as leptons and baryons.
The justification for this comes from considering the electromagnetic interaction with other matters. Let us consider,
in addition to gravitational matter ($G_i$),
a system consisting of the EM field $E_i$ and another field $B_{ijk}$.
If we are interested in electromagnetic/hard-matter interaction, we can write the tensor of matter M symbolically as:
\begin{equation}
\label{f20}
M=\Sigma (E^{{\,}{\prime}}+B^{{\,}{\prime}}+E^{\,2}+B^{\,2}+EB)
\end{equation}
Above, $E^{{\,}{\prime}}$ represents a covariant derivative of $E_i$ ($E_{i;j}$). Then, per eq. (\ref{f2}),
the Lagrangian is proportional to the square of M and thus can be written symbolically as:
\begin{equation}
\label{f21}
L=L_{em}(E^{\,\prime})+L_{mat}(B^{\,\prime},B^2)+L_{em\_int}
\end{equation}
The interaction term $L_{em\_int}$ must be proportional to $B^2$. This statement comes from the fact that when we write the system of equations
for the electromagnetic field and the tensor $B$, the equations for the tensor B must vanish, in the case when no hard matter is present ($B=0$).
If  the Lagrangian of hard-matter and the EM field has a term like $B^{\,\prime}E^{\,\prime}$, the equations for the tensor $B$ when $B=0$
yield 3-index equations of a type $E^{\,\prime\prime}=0$. That would mean that $E=0$.  Thus, we can write the following:
\begin{equation}
\label{f21a}
L_{em\_int}=E^{\, \prime}B^{\, 2}
\end{equation}
We must remember that the vector $E$ always comes together with the tensor Levi-Civita (${\epsilon}_{ijkl})$. 
An example of such Lagrangianian terms can be:
\begin{eqnarray}
\label{f21b}
&&L_{em\_int}=[B_{ijk}{B_l}^{ij}-B_{ijl}{B_k}^{ij}]{\epsilon}^{klmn}E_{m;n}=\nonumber\\
&&=-{[B_{ijk}{B_l}^{ij}-B_{ijl}{B_k}^{ij}]}_{;n}{\epsilon}^{klmn}E_m ={X^{mn;}}_nE_m
\end{eqnarray}
Above, we have used the rule of variational analysis similar to the one of partial integration. We also introduced a tensor $X_{mn}$ as
\begin{equation}
\label{f21c}
X^{mn}=-[B_{ijk}{B_l}^{ij}-B_{ijl}{B_k}^{ij}]{\epsilon}^{klmn}
\end{equation}

Of course, $X^m={X^{mn}}_{;n}$ serves as an electrical current associated with matter $B_{ijk}$. 

It is interesting to note, that if we replace the tensor Levi-Civita with the imaginary unit ($i=\sqrt{-1}$), as it is sometimes done,
the expression above looks very much like an expression for the electrical current in quantum mechanics.

The expression (\ref{f21c}) for $X^{mn}$ is one of several that could be derived from the symbolic expression (\ref{f21a}).
However, the expression (\ref{f21c}) is the only one that has anti-symmetrical properties.
In this case, the diversion of the vector $X^m$ is zero ($X^m_{;m}=0$), which comes directly from the anti-symmetry of the tensor $X^{mn}$ 
and the definition of the current $X^m (X^m=X^{mn}_{;n}$).

Similar formulas and conclusions can be made for the tensor $C_{ijk}$. If we mark ${X^+}_m$ as the EM current for the matter $B_{ijk}$
and ${X^-}_m$ as the current for the matter $C_{ijk}$, we must conclude that not only the total charge is conserved
${({X^+}_i+{X^-}_i)}^{;i}=0$, but that each of their corresponding parts - $X^+$ and $X^-$ - are also conserved. In other words, 
both electron (lepton) charge and proton (baryon) charge are conserved.

It is not difficult to see that both the expression (\ref{f21c}), written for the fully
symmetrical tensor $G_{ijk}$, and the corresponding electrical current will vanish due to the presence of the fully 
anti-symmetrical tensor Levi-Civita (${\epsilon}_{ijkl})$. 

The above reasoning is the justification for our attempt to identify tensor $G_{ijk}$ with the neutron and $B_{ijk}$, $C_{ijk}$
with charged particles - the proton and electron. The choice of $B_{ijk}$ for the proton is based on the fact that both tensors,
$G_{ijk}$ and $B_{ijk}$, are derived from the same symmetrical in the low indices part of tensor  $P^i_{jk}$. 

By associating tensors $G_{ijk}$, $B_{ijk}$ and $C_{ijk}$ with elementary particles,
we radically depart from the well established spinor representation.
The hope here is that 16 independent functions in each tensor are capable of describing the experimentally observable effects
of elementary particles that are presently described by 1/2-spinor.

The other terms that correspond to EM-hard-matter interaction can be symbolically written as $C^2E^2$.
It is not clear if such terms can be present in the Lagrangian. But if they can exist, the Maxwell equations will
have an additional term(s) proportional to the vector potential (e.g. $E_iB_{klm}B^{klm}$). 

We still have to address the nature of the other two vector fields, $B_i$ and $C_i$.
We showed before that a combination of only two vectors, $G_i$ (gravitation) and $B_i$ - in addition to the metric $g_{ij}$ - cannot exist.
The definition of the metric tensor $g_{ij}$, eq. (\ref{f8}), prevents this from occurring. 
In other words, the matter represented by the vector $B_i$ 
cannot propagate through the space outside the immediate vicinity of hard matter. On the other hand, the vector $C_i$, has no such restrictions
if we chose constants of eq. (\ref{f8}) to be (-18, 12, 1). By analogy with the electromagnetic field, it would be logical to 
assume that these vector fields would be the media, through which the hard particles, $G_{ijk},\,B_{ijk},\,C_{ijk}$, exchange their 
energy over the distance beyond their immediate reach. Could these fields be responsible for an electro-weak and nuclear interactions?

\vskip 3em
{\bf \underline{Unconstrained Metric}}
\vskip 1em

In the end, we have to mention another formulation of the theory for a tensor potential description of matter.
In this formulation, we start a theory with the assumption that both $P^i_{jk}$ and $g_{ij}$ are independent variables of the theory.
In this case, the Lagrangian multipliers $T^{ij}$ in eq, (\ref{f29a}) would not be present and the law of conservation would be exact without
any additional requirements. The law of conservation would be guaranteed by the fact that the Lagrangian is proportional to the square
of the tensor M or $g^{ij}$. 
Also, we would have more freedom in choosing, among the four vectors $G_i$, $B_i$, $C_i$ and $E_i$, the vector representing the gravitational field.
For example, we could have chosen the vector $E_i$ (or fully anti-symmetrical $P_{ijk}=({\epsilon} _{ijkm}E^m)/6)$ as a gravitational field.
Eveything that we said about gravitational fields and electromagnetic-gravitational interactions would hold true. 
Contrary to the case of the "constrained" theory described in this article, where matter defined by only two vectors $G_i$, $B_i$ could not exit,
there are no such restrictions in this unconstrained formulation.
In fact, the basic vector fields $G_i$, $B_i$, $C_i$ and $E_i$ can be derived from $P^i_{jk}$ differently to the way
it has been done previously in this article. For example, the electromagnetic field could be defined as $E_i=P^m_{mi}$ 
and the gravitational field as $G^i={P^i}_{jk}g^{jk}$.

The equations of motion in the case of a constant potential $P^i_{jk}$ becomes algebraic with respect to $P^i_{jk}$ and $g_{ij}$. If a solution exists
then it corresponds to flat space.

The electromagnetic field still can not exist by itself if the Lagrangian has Maxwell's form.

Since the Lagrangian is proportional to at least the square of $P^i_{jk}$, the Mach priciple still holds. This is due to the fact that
in the absence of matter ($P^i_{jk}=0$), all equations vanish and the metric tensor remains undetermined.

While the above mentioned factors of unconstrained theory should be viewed as advantages, there are drawbacks to this approach.
First, we are walking away from a singular entity ($P^{\,i}_{jk}$) description of matter and space.
Secondly, the metric field is defined only up to the constant factor. Thus, we can not say that a larger field corresponds 
to a larger space metric. Additional relations between a metric and a tensor potential have to be postulated to fix that problem.

Philosophically, having $g_{ij}$ as an independent variable suggests that space is an independent entity.
In other words, with an independent $g_{ij}$ we gain significant freedom in describing matter, however, we lose philosophical elegance.
The latter factor can not outweigh the former, and the unconstrained theory can not be dismissed a priori without any investigation.

\vskip 3em
{\bf \underline{Conclusion}}
\vskip 1em

The theory proposed in this article is just a sketch, an idea, a thought. If there is any truth to such description of matter,
it would require significant work to be done before the theory takes the shape of THE theory. This work should be done both
in the field of mathematics and in the field of physics. 

Mathematically, we have to prove that at least for some Lagrangians our postulate T=0 does not contradict the equations of motion
(eqs. \ref{f29},  \ref{f29a},  \ref{f29b}). What do the four Noether identities yield regarding the tensor energy-momentum $T_{ij}$;
do they lead to ${T^{ij}}_{;j}=0$?  Are the elementary particles (tensor $S_{ijk}$, $C_{ijk}$, $B_{ijk}$) true singularities
or finite deviations of the background field? Do the equations that define the behavior of these tensors allow quantum states?

We need to derive Enstein's postulate that a "small" macroscopic body, as well as light, travels along geodesic lines of metric $g{ij}$.
And if the postulate is true (which it seems to be for celestial bodies), how "small" can the body be? Can it be as small as a molecule
or an atom? Does it have to be electrically neutral and is it also true for ions? 
And what effect does the background gravitational field have on the motion of such bodies?

Physically, we have to identify the constants in the equation that defines the metric tensor $g_{ij}$ (eq. \ref{f25}). 
We have to answer the question of the background potential ${{Po}^i}_{jk}$. Is it built only out of the gravitational field
or does it include some other fields? We have to definitively identify which two (out of four) vector fields correspond to
gravitation and electromagnetism. As we mentioned before there are other possibilities (aside of $G_i$ and $E_i$).
For example, $G_i$ and $C_i$. What is the meaning of the other two vector fields?

And most importantly, we have to identify the Lagrangian that describes well both macro physics (gravitation, electromagnetism)
and micro physics (elementary particles and nuclear interactions).

\vskip 3em
{\bf\underline{\underline{Appendix A}}}
\vskip 1em

In this appendix we give a full derivation of the equations of motion associated with the variation of the action A with respect to $g_{ij}$.

\begin{equation}
A=\int\nolimits L(P^i_{jk;l}, P^i_{jk}, g_{ij}){\,}{\sqrt g}{\,}d^4x
\end{equation}
Above, $g$ is a minus determinant of $g_{ij}\,(g=-det(g_{ij})\,)$.

The Lagrangian L depends on the tensor $g_{ij}$ in two ways: a) algebraically, including the square root of the $g$ term and b)
through the covariant derivative  of $P^i_{jk}$, by means of Christoffel symbols ${\gamma}^i_{jk}$. 

\begin{eqnarray}
\label{f73}
&&{\delta}A={\delta}{\int\nolimits L({P^i}_{jk;l}, {P^i}_{jk}, g_{ij}){\,}{\sqrt g}{\,}d^4x}
={\int\nolimits} {\delta}(L{\sqrt g}){\,} d^4x= \nonumber\\
&&={\int\nolimits}{\,} {\sqrt{g}}{\,}d^4x  [({\frac {{\partial}L} {{\partial}g_{ij}}}+1/2Lg^{ij}){\delta}g_{ij}+
({\frac {{\partial}L}{{\partial}{{P^i}_{jk;l}}}){\delta}{P^i}_{jk;l}}]
\end{eqnarray}
Let us introduce the tensor ${L_i}^{jkl}={\partial}L/{\partial}{P^i}_{jk;l}$. The last term in (\ref{f73}) then can be written as:
\begin{eqnarray}
\label{f73a}
&&\int\nolimits d^4x{\sqrt{g}} {L_i}^{jkl} {\delta}{P^i}_{jk;l}=\nonumber\\
&&\int\nolimits d^4x{\sqrt{g}} {L_i}^{jkl}{\delta}[{\frac {{\partial}{P^i}_{jk}} {{\partial}{x^l}}}
+{\gamma}^i_{ls}{P^s}_{jk}-{\gamma}^s_{lj} {P^i}_{sk}-{\gamma}^s_{lk} {P^i}_{js}]= \nonumber\\
&&\int\nolimits d^4x{\sqrt{g}} {L_i}^{jkl}{P^s}_{jk}{\delta}{{\gamma}^i}_{ls}-{L_i}^{jkl}{P^i}_{sk}{\delta}{{\gamma}^s}_{lj}-
L_i^{jkl}P^i_{js}{\delta}{{\gamma}^s_{lk}}= \nonumber\\
&&\int\nolimits d^4x{\sqrt{g}} [{L_i}^{mnk}P^j_{mn}-L_m^{jnk}P^m_{in}-L_m^{njk}P^m_{in}]{\delta}{\gamma}^i_{jk}=
\end{eqnarray}
To be compatible to the fact that the Christoffel symbols ${\gamma}^i_{jk}$ are symmetrical by $j,k$, the last expression needs to be symmetrized
by indices $j,k$.

\begin{eqnarray}
\label{f73b}
&&=\int\nolimits d^4x{\sqrt{g}} {\frac{1}{2}}[({L_i}^{mnk}P^j_{mn}-{L_m}^{jnk}P^m_{in}-{L_m}^{njk}P^m_{in}) \nonumber\\
&&+({L_i}^{mnj}P^k_{mn}-{L_m}^{knj}P^m_{in}-{L_m}^{nkj}P^m_{in})]{\delta}{\gamma}^i_{jk}= \nonumber\\
&&\int\nolimits d^4x{\,}{\sqrt{g}}{\,}{\frac{1}{2}}J^{jk}_i{\delta}{\gamma}^i_{jk}
\end{eqnarray}

To shorten the formula, we have introduced a new symmetrical in low indices tensor $J^{jk}_i$ for the  terms in square brakets.
We can now replace the Christoffel
symbols with their expression through the matrix tensor $g_{ij}$. The variation of the Christoffel symbols consists of two terms:
${\delta}{\gamma}^i_{jk}={\delta}(g^{is} {\gamma}_{sjk})={\gamma}_{sjk}{\delta}g^{si}+g^{is}{\delta}{\gamma}_{sjk}$. 
The first term (${\gamma}_{sjk}{\delta}g^{si}$), which is proportional to the ${\gamma}$, can be omitted due to the fact
that it must be set to zero when we replace all partial derivatives with the covariant ones.
With that in mind, the expression above could be written as:
\begin{eqnarray}
\label{f73c}
&&\int\nolimits d^4x{\sqrt{g}}{\,}{\frac{1}{2}} J^{jk}_i{\delta}{\gamma}^i_{jk}=\int\nolimits d^4x{\sqrt{g}}
{\frac{1}{4}}J^{jk}_i{\delta}[g^{is}(g_{sj,k}+g_{sk,j}-g_{jk,s})] \nonumber\\
&&=\int\nolimits d^4x{\sqrt{g}} {\frac{1}{4}}J^{jk}_ig^{is}({\delta}g_{sj,k}+{\delta}g_{sk,j}-{\delta}g_{jk,s})=\\
&&\int\nolimits d^4x{\sqrt{g}} {\frac{1}{4}}[-{({J^{jk}_ig^{is}})}_{,k}{\delta}g_{sj}-{({J^{jk}_ig^{is}}})_{,j}{\delta}g_{sk}
+{({J^{jk}_ig^{is}})}_{,s}{\delta}g_{jk}] \nonumber
\end{eqnarray}
In eq. (\ref{f73c}), we have used the rule of variational analysis which is similar to the rule of partial integration. 
We now rename the indices in a such way that
in the end we are going to have only one ${\delta}g_{ij}$ term:
\begin{eqnarray}
\label{f73d}
&&\int\nolimits d^4x{\sqrt{g}} {\frac{1}{4}} {[-J^{ijs}-J^{jis}+J^{sij}]}_{,s}{\delta}g_{sk}
\end{eqnarray}
Above, $J^{ijk}=g^{im}J_m^{jk}$.

Combining expressions (\ref{f73d}) and ({\ref{f73b}), we can write the final expression for the variation of the action A:
\begin{eqnarray}
&&{\delta}A= \nonumber\\
&&\int\nolimits d^4x{\sqrt{g}}[({\frac {{\partial}L} {{\partial}g_{ij}}}+1/2Lg^{ij})+
 {\frac{1}{4}} ({-J^{ijs}-J^{jis}+J^{sij})}_{,s}]{\delta}g_{sk}
\end{eqnarray}

Thus, the equation of motion associated with the variation of the action A with respect to the $g_{ij}$ will be:

\begin{eqnarray}
&&{\frac {{\partial}L} {{\partial}g_{ij}}}+{\frac{1}{2}}{\,} Lg^{ij}+{\frac{1}{4}}{\,}{[-J^{ijs}-J^{jis}+J^{sij}]}_{;s}=0  \\
&&J^{ijs}=g^{sr}[({L_r}^{mni}P^j_{mn}-{L_m}^{jni}P^m_{rn}-{L_m}^{nji}P^m_{rn}) \nonumber\\
&&+({L_s}^{mnj}P^i_{mn}-{L_m}^{inj}P^m_{sn}-{L_m}^{nij}P^m_{sn})] \\
&& {L_i}^{jkl}={\partial}L/{\partial}P^i_{jk;l}
\end{eqnarray}

\vskip 3em
{\bf\underline{\underline{Appendix B}}}
\vskip 1em

In this appendix we will give a detailed expression of the tensor potential $P_{ijk}$ through simple tensorial quantities.
First, we can separate $P_{ijk}$ into fully symmetrical $G_{ijk}$ and two anti-symmetrical tensors ($B_{ijk}$ and $C_{ijk}$)
with respect to 2nd and 3rd indices.

\begin{equation}
\label{f82}
P_{ijk}=(P_{ijk}+P_{ikj})/2+(P_{ijk}-P_{ikj})/2={\bar P}_{ijk}+{\hat P}_{ijk}
\end{equation}
Above, ${\bar P}_{ijk}$ and ${\hat P}_{ijk}$ are symmetrical and anti-symmetrical in 2-3 indices tensors, respectively.
We now introduce a fully symmetrical tensor ${\bar S}_{ijk}=({\bar P}_{ijk}+{\bar P}_{jki}+{\bar P}_{kij})/3$. Using ${\bar S}_{ijk}$
we can rewrite the tensor ${\bar P}_{ijk}$ in the following form:
\begin{eqnarray}
\label{f83}
&&{\bar P}_{ijk}=[{\bar P}_{ijk}-({\bar P}_{ijk}+{\bar P}_{jki}+{\bar P}_{kij})/3]+{\bar S}_{ijk}=\nonumber\\
&&({\bar P}_{ijk}-{\bar P}_{jki})/3+({\bar P}_{ijk}-{\bar P}_{kij})/3+{\bar S}_{ijk}=\nonumber\\
&&({\bar P}_{ikj}-{\bar P}_{jki})/3+({\bar P}_{ijk}-{\bar P}_{kji})/3+{\bar S}_{ijk}=\nonumber\\
&&{\bar B}_{kij}+{\bar B}_{jik}+{\bar S}_{ijk}
\end{eqnarray}
Above, ${\bar B}_{ijk}=({\bar P}_{jik}-{\bar P}_{kij})/3$ is an anti-symmetrical in 2nd-3rd indices tensor.
Since ${\bar B}_{ijk}$ is defined through the symmetrical in 2nd-3rd indices tensor ${\bar P}_{ijk}$, it satisfies the following equation:
\begin{equation}
\label{f84}
{\bar B}_{ijk}{\epsilon}^{ijkl}=0
\end{equation}
In the equation above, ${\epsilon}^{ijkl}$ is the fully anti-symmetrical tensor of Levi-Civita.

Due to their symmetries, each of the tensors ${\bar S}_{ijk}$, ${\bar B}_{ijk}$ has only one vector produced by a contraction of indices.
This allows us to split each tensor into a traceless tensor and a vector. Tensor ${\hat P}_{ijk}$, however, has two 
different vectors that can be formed from it.

\begin{eqnarray}
\label{f85}
&&{\bar S}_{ijk}=G_{ijk}+1/6(G_ig_{jk}+G_jg_{ik}+G_kg_{ij})\nonumber\\
&&{\bar B}_{ijk}=B_{ijk}+1/3(B_kg_{ij}-B_jg_{ik})\nonumber\\
&&{\hat P}_{ijk}=C_{ijk}+1/3(C_kg_{ij}-C_jg_{ik})+ 1/6{\epsilon}_{ijkm}E^m
\end{eqnarray}
Combining (\ref{f82}, \ref{f83}, and \ref{f85}) we will get:
\begin{eqnarray}
\label{f86}
&&P_{ijk}=[G_{ijk}+1/6(G_ig_{jk}+G_kg_{ij}+G_jg_{ki})] + \nonumber\\
&&[(B_{jik}+B_{kij})+1/3(B_jg_{ik}+B_kg_{ij}-2B_ig_{jk})]+ \nonumber\\
&&[C_{ijk}+1/3(C_kg_{ij}-C_jg_{ik})]+ 1/6{\epsilon}_{ijkm}E^m
\end{eqnarray}
This representation is fully reversible. It allows us to express all individual components through the original tensor $P^i_{jk}$.
\begin{eqnarray}
\label{f87}
&&G_i=(P^k_{ki}+P^k_{ik}+P^m_{nl}g^{nl}g_{mi})/3 \nonumber\\
&&B_i=(P^k_{ik}+P^k_{ki}-2P_{ikl}g^{kl})/6\nonumber\\
&&C_i=(P^k_{ki}-P^k_{ik})/2 \nonumber\\
&&E_i={{\epsilon}_i}^{jkm}P_{jkm}\nonumber\\
&&G_{ijk}=(P_{ijk}+P_{jki}+P_{kij}+P_{ikj}+P_{kji}+P_{jik})/6\nonumber\\
&&-(P^k_{ki}+P^k_{ik}+P^m_{nl}g^{nl}g_{mi})/3 \nonumber\\
&&B_{ijk}=(P_{jik}+P_{jik}-P_{kij}-P_{kji})/6-(P^k_{ik}+P^k_{ki}-2P_{ikl}g^{kl})/6 \nonumber\\
&&C_{ijk}=(P_{ijk}-P_{ikj})/2-(P^k_{ki}-P^k_{ik})/2
\end{eqnarray}

\vskip 3em
{\bf\underline{\underline{Appendix C}}}
\vskip 1em
In this appendix we will give a step-by-step derivation of the Euler equations of motion for the Lagrangian $L=L(G_{i;j}, G_i, g_{ij})$
with respect to the tensor $g_{ij}$. The process is very similar to the one done in Appendix A.
\begin{eqnarray}
\label{f88}
&&{\delta}A={\delta}\int\nolimits L(G_{i;j},G_i,g_{ij}){\,}\sqrt{g}d^4x =\nonumber\\
&&{\delta}\int\nolimits d^4x \sqrt{g}{\,} {\delta}L(G_{i;j},G_i,g_{ij})=\nonumber\\
&&\int\nolimits d4x {\delta}[ \sqrt{g}{\,}L(G_{i;j},G_i,g_{ij})]=\nonumber\\
&&\int\nolimits d^4x \sqrt{g}{\,}\{ [{\partial}L/{\partial}g_{ij}+1/2Lg^{ij}]{\,}{\delta}g_{ij}+
{\partial}L/{\partial}G_{i;j}\}{\,}{\delta}G_{i;j}\nonumber\\
&&\int\nolimits d^4x \sqrt{g}{\,}\{ [{\partial}L/{\partial}g_{ij}+1/2Lg^{ij}]{\,}{\delta}g_{ij}+
{\partial}L/{\partial}G_{i;j}{\,}{\delta}{\gamma}^p_{ij}G_p\} \nonumber\\
&&\int\nolimits d^4x \sqrt{g}{\,} \{[{\partial}L/{\partial}g_{ij}+1/2Lg^{ij}]{\,}{\delta}g_{ij}-\nonumber\\
&&1/2[{\partial}L/{\partial}G_{i;j}+{\partial}L/{\partial}G_{j;i}]{\,}G^p{\delta}{\gamma}_{pij}\}=\nonumber\\
&&\int\nolimits d^4x \sqrt{g}{\,}\{ [{\partial}L/{\partial}g_{ij}+1/2Lg^{ij}]{\,}{\delta}g_{ij}+\nonumber\\
&&L^{ij}{\,}G^p(-1/2)({\delta}g_{ip,j}+{\delta}g_{jp,i}-{\delta}g_{ij,p})\}
\end{eqnarray}
Above, $L^{ij}=1/2[{\partial}L/{\partial}G_{i;j}+{\partial}L/{\partial}G_{j;i}]$.

We symmetrized the expression for $L^{ij}$ 
due to the fact that ${\delta}{\gamma}_{pij}$ is symmetrical in the 2nd-3rd indices.

We also omitted the term containing $({\delta}g^{pm}){\gamma}_{mij}G_p$ because it vanishes during trasition from the partial
derivitives to the covariant ones.
Integrating terms like ${\delta}g_{ip,j}$ by partial integration, we continue the expression above as:
\begin{eqnarray}
\label{f89}
&&\int\nolimits d^4x \sqrt{g}{\,} [{\partial}L/{\partial}g_{ij}+1/2Lg^{ij}]{\,}{\delta}g_{ij}+\nonumber\\
&&(1/2){\{}{[L^{ij}{\,}G^p]}_{,j}{\delta}g_{ip}+{[L^{ij}{\,}G^p]}_{,i}{\delta}g_{jp}-{[L^{ij}{\,}G^p]}_{,p}{\delta}g_{ij} {\}}
\end{eqnarray}
We can change indices in the last three terms in such a way that we only have ${\delta}g_{ij}$:
\begin{eqnarray}
\label{f90}
&&\int\nolimits d^4x \sqrt{g} {\,} {\{}[{\partial}L/{\partial}g_{ij}+1/2Lg^{ij}]{\,}{\delta}g_{ij}+\nonumber\\
&&(1/2)[{(L^{ip}G^j )}_{,p}+{(L^{pj}G^i )}_{,p}-{(L^{ij}{\,}G^p )}_{,p}]{\}}{\delta}g_{ij}
\end{eqnarray}

Commas, as partial derivatives, can be replaced with semi-columns (;) for covariant derivatives, and
the equation ${\delta}A/{\delta}g_{ij}=0$ yields:
\begin{eqnarray}
\label{f91}
&&[{\partial}L/{\partial}g_{ij}+1/2Lg^{ij}]+\nonumber\\
&&(1/2)[{(L^{ip}G^j)}_{;p}+{(L^{pj}G^i)}_{;p}-{(L^{ij}{\,}G^p)}_{;p}]=0
\end{eqnarray}

We can now consider some particular Lagrangians as a function of $G_i$. 
Thus, for $L$ given by eq. (\ref{f9}) $L=G_{i;j}G_{k;l}g^{ik}g^{jl}-{(G_{i;j}g^{ij})}^2$, we will get:
\begin{equation}
\label{f92}
L^{ij}=[G^{i;j}+G^{j;i}]-[2{G^k}_{;k}g^{ij}]
\end{equation}
The equation ${\delta}A/{\delta}g_{ij}=0$ with all indices lowered down will have this expression:
\begin{eqnarray}
\label{f93}
&&[-G_{i;k}G_{l;j}g^{kl}-G_{j;k}G_{l;i}g^{kl}+1/2(G_{k;l}G_{m;n}g^{km}g^{ln})]-\nonumber\\
&&[-{G^k}_{;k}(G_{i;j}+G_{j;i})+1/2{({G^k}_{;k})}^2g_{ij}]+ \nonumber\\
&&1/2[{(G_{j;p}+G_{p;j})G_i}^{;p}+{(G_{i;p}+G_{p;i})G_j}^{;p}-{(G_{j;i}+G_{i;j})G_p}^{;p}]-\nonumber\\
&&[{G^k}_{;k}(G_{i;j}+G_{j;i})-{({G^k}_{;k}g_{ij}G_p)}^{;p}]
\end{eqnarray}
Contracting it by indices $i-j$ we will have this expression for the law of conservation:
\begin{eqnarray}
\label{f95}
&&{J^p}_{;p}=0\quad where \nonumber\\
&&J_p=G_{i;p}G^i+G_{p;i}G^i+{G^i}_{;i}G_p
\end{eqnarray}
In a more general case, when the Lagrangian is given by an expression:
\begin{equation}
\label{f96}
L=k_1G_{i;j}G^{i;j}+k_2G_{i;j}G^{j;i}+k_3{({G_{i;}}^i)}^2+k_4{({G_{i;}}^i)}G^{\,k}G_k
\end{equation}
equation ${\delta}A/{\delta}g_{ij}=0$ takes this form:
\begin{eqnarray}
\label{f97}
&&k_1[-G_{i;k}G_{j;l}g^{kl}-G_{k;i}G_{l;j}g^{kl}+1/2(G_{k;l}G^{k;l})g_{ij}]-\nonumber\\
&&k_2[-G_{i;k}G_{l;j}g^{kl}-G_{j;k}G_{l;i}g^{kl}+1/2(G_{k;l}G^{l;k})g_{ij}]-\nonumber\\
&&k_3[-{G^k}_{;k}(G_{i;j}+G_{j;i})+1/2{({G^k}_{;k})}^2g_{ij}]+ \nonumber\\
&&k_4[-{({G_{k;}}^k)}G_iG_j+1/2g_{ij}{({G_{k;}}^k)}G^mG_m]+ \nonumber\\
&&(k_1+k_2)/2 {[(G_{j;p}+G_{p;j})G_i+(G_{i;p}+G_{p;i})G_j-(G_{j;i}+G_{i;j})G_p]}^{;p}+\nonumber\\
&&k_3{[{G^k}_{;k}G_{i;j}+{G^k}_{;k}G_{j;i}-{G^k}_{;k}g_{ij}G_p]}^{;p}+ \nonumber\\
&&k_4/2[{(G^kG_kG_i)}_{;j}+{(G^kG_kG_j)}_{;i}-g_{ij}{(G^mG_mG^k)}_{;k}=0
\end{eqnarray}
with the law of conservation given by the formula below:
\begin{eqnarray}
\label{f98}
&&{J^p}_{;p}=0\quad where \nonumber\\
&&J_p=(k_1+k_2)(G_{i;p}G^i+G_{p;i}G^i)-\nonumber\\
&&(k_1+k_2+2k_3){G^i}_{;i}G_p-k_4G^iG_iG_p
\end{eqnarray}

\end{document}